\documentclass[aps,prb,twocolumn,superscriptaddress,floatfix,longbibliography]{revtex4-2}

\usepackage[utf8]{inputenc}
\usepackage{graphicx}
\usepackage{braket}
\usepackage{bm}
\usepackage{upgreek}
\usepackage{mathtools}
\usepackage{array}
\usepackage{physics}

\usepackage{hyperref}

\begin{document}
\title{Impact of the valence band on Rydberg excitons in cuprous oxide quantum wells}

\author{Niklas Scheuler}
\author{Jörg Main}
\email[Email: ]{main@itp1.uni-stuttgart.de}
\author{Patric Rommel}
\author{Frieder Pfeiffer}
\affiliation{Institut für Theoretische Physik I, Universität
  Stuttgart, 70550 Stuttgart, Germany}
\author{Stefan Scheel}
\author{Pavel A. Belov}
\affiliation{Institut für Physik, Universität Rostock,
  Albert-Einstein-Straße 23-24, 18059 Rostock, Germany}

\date{\today}

\begin{abstract}
The complex valence band structure of bulk cuprous oxide necessitates
going beyond the parabolic approximation to precisely estimate exciton
binding energies.  The same is true for excitons in cuprous oxide
quantum wells, for which many effects have been obtained so far only
qualitatively within a hydrogenlike two-band model.  Here, we derive
the complete Hamiltonian for excitons in cuprous oxide quantum wells
based on the Luttinger-Kohn model, taking into account the full
complex valence band structure.  Symmetry properties of the system are
discussed.  Numerical results based on the diagonalization of the
Hamiltonian using B-spline functions reveal the energy shifts and the
lifting of degeneracies due to the nondiagonal coupling terms of the
complex valence band.  The relative oscillator strengths of the
excitonic transitions induced by circularly polarized light are also
calculated.
\end{abstract}

\maketitle

\section{Introduction}
\label{sec:intro}
Electron-hole bound states in bulk cuprous oxide are characterized by
large binding energies, allowing for high-resolution measurements of
the excitonic transitions of yellow and green Rydberg
series~\cite{Kaz14,Schweiner16b,Heck17a}.
This paves the way for excitonic devices~\cite{Butov2017},
sensors~\cite{Meyer2020,Chukeev2024} and potentially room-temperature
applications~\cite{Grosso2009,Andreakou2014,Butov2017}.
Due to strong spin-orbit coupling, an accurate theoretical description
of exciton spectra in bulk Cu$_{2}$O requires taking into account the
complex valence band structure that goes beyond the simple parabolic
band approximation.
Effects of the band structure on exciton energies have been
investigated recently both theoretically and experimentally
\cite{Thewes15,Schoene16,Schweiner16a,Schweiner17a,Rommel18,Heck18,Heck25}.
These works have significantly advanced our understanding of the
exciton properties in bulk Cu$_{2}$O, in comparison to the pioneering
studies of exciton energies and
absorption~\cite{Gross1,Gross2,Uihlein1981}.

Similar to other types of semiconductors such as GaAs~\cite{Davies1997,Alferov}
or CdSe~\cite{Efros2000,Efros2021}, a greater tunability of
electronic and optical properties of quasiparticles in Cu$_{2}$O can
be achieved by fabrication of
heterostructures~\cite{Lynch2021,DeLange2023}.
The first films of cuprous oxide sandwiched between sapphire
substrates and thus realizing the weak quantum confinement of lower
exciton states have already been obtained~\cite{Naka2018,NakaPRL}.
When the size of a particular exciton state is comparable to the
thickness of the film, the confinement effect dominates.
Therefore, for very thin Cu$_{2}$O films or for Rydberg exciton
states, the regime of strong confinement is reached, and the details of
the complex band structure play the role of a small perturbation. As a
result, the parabolic approximation of the bands in the vicinity of
the $\Gamma$ point is a reasonable choice. However, for weak
confinement the effects due to features of the band structure are
comparable to the quantum-confinement effects. Thus, the full Cu$_{2}$O
band structure has to be explicitly taken into account.

The quantum confinement appearing due to the band offsets at the
heterojunction can, in its simplest form, be modeled by the quantum
well (QW) potential~\cite{Bastard1982}.
Excitons in QWs exhibit a large variety of interesting phenomena and
effects~\cite{ivchenko,Khramtsov2016}.
The theory of excitons and their radiative properties in GaAs-based
QWs is well developed~\cite{Andreani1991,Davies1997}, though for cuprous oxide 
finite-sized crystals the energy spectrum as well as the relative
oscillator strengths have been obtained only very recently
\cite{Belov2024,Kuehner2025}.
These works significantly differ from a simplified consideration in
Refs.~\cite{Ziemkiewicz2021a,Ziemkiewicz2021b} by studying the
spectrum as a function of the QW width.
A dependence of the exciton energy levels on the width of a QW allows one
to study a crossover from weak confinement in the bulk to nearly
two-dimensional excitons in case of strong confinement.
Besides bound states, above the scattering threshold electron-hole resonances occur.
Their spectral lines are usually additionally broadened due to coupling to the continuum.
For GaAs-based QWs, the energies and linewidths of the electron-hole
resonances have been estimated using Feshbach theory~\cite{Feshbach1958} in
Refs.~\cite{Broido1988,Pasquarello1991}.
The resonant broadenings can also be computed by both the
stabilization and the complex-coordinate-rotation method, as it was
done for GaAs-based \cite{BelResQW} and Cu$_{2}$O-based QWs~\cite{Scheuler2024}.
When varying the confinement strength, so-called bound states in
the continuum (BICs), i.e., long-lived Rydberg states of the confined
electron-hole pairs, appear in the continuum background~\cite{Aslanidis2025}.
The BICs are uncoupled from the continuum of lower quantum-confinement
subbands and thus have zero linewidths~\cite{Hsu2016,Happ2025}.
All these effects can be successfully studied numerically by expanding
the electron-hole wave function over a basis of B-splines~\cite{DeBoor,Bachau2001}.
The wave functions in coordinate space, computed using a B-spline
representation, clearly illustrate, e.g., the quenching of states in
the strong confinement regime, the localized nature of BICs and,
furthermore, allow for the estimation of oscillator strengths.

In Refs.~\cite{Scheuler2024,Belov2024,Kuehner2025}, the energy levels of 
excitons and electron-hole resonances in finite-sized cuprous oxide crystals 
have been calculated using the two-band model with simple parabolic
dispersions of the bands.
Although these works shed light on the crossover from the weak 
to the strong confinement regime as well as on the radiative properties of exciton
states of different symmetries, the exciton states were studied within
the hydrogenlike model, confined over one dimension.
The complex structure of the Cu$_2$O valence subbands near the
$\Gamma$ point has not been considered so far.
Taking into account the complex valence band dispersion leads to an
increase in the number of coupled differential equations corresponding
to different projections of the orbital momemtum and spin.
This makes the numerical treatment of the problem much more difficult.

The transition from bulk to a monolayer affects exciton states in many ways. 
Apart from restricting the motion of electron and hole along one dimension,
it also leads to a suppression of the exciton kinetic energy,
a change of the band structure and a distortion of the Coulomb interaction.
The reduction of spatial dimension from three to two dimensions
reduces the kinetic energy, thereby increasing the exciton binding
energy from $E^{3D}_{b}=\text{Ry}/n^2$ to $E^{2D}_{b}=\text{Ry}/(n-1/2)^2$,
where $\text{Ry}$ is the Rydberg energy and $n=1,2,\ldots$ is the
principal quantum number~\cite{ivchenko,Kezerashvili2019}.
The band structure also depends on the geometry of the crystal
structure: the effective mass parameters in the kinetic terms are
modified during a crossover from bulk to a monolayer.
Moreover, this crossover affects the potential energy of the
electron-hole interaction.
The dielectric contrast between the thin film and the environment
(vacuum or a substrate) distorts the pure Coulomb interaction.
For thin films, the resulting screening effect can be described by 
the Rytova-Keldysh potential~\cite{Rytova,Keldysh}.
Its impact on exciton states is well known for transition metal dichalcogenide
monolayers~\cite{Chernikov2014,Cavalcante2018,Kezerashvili2020}, phosphorene
monolayers~\cite{Brunetti2019}, and has been studied for Dirac
materials~\cite{Leppenen2020} and cuprous oxide thin films~\cite{Belov2024}.

The aim of this article is to study the impact of the complex valence
band on excitons in cuprous oxide QWs.
Besides the quantization over the confinement direction, we consider
the Luttinger Hamiltonian~\cite{Luttinger56} or, equivalently,
the Suzuki-Hensel Hamiltonian~\cite{Suzuki1974}
for the hole dispersion in the QW together with a purely parabolic
dispersion for the electron.
Such a hole dispersion significantly complicates the Schr\"{o}dinger
equation for excitons in cuprous oxide QWs~\cite{Belov2024,Kuehner2025}.
Moreover, it breaks some exact symmetries of the pure parabolic
dispersion, in particular, the rotational invariance around the axis
normal to the QW plane, causing the angular momentum quantum number
$m$ to be only an approximate quantum number.
Nevertheless, using the B-spline representation of the wave function,
we can calculate the matrix elements of all involved operators and solve
the eigenvalue problem to obtain the energy levels as a function of
the QW width.
We study effects of the complex valence band, originating from
coupling of states with different values of $m$.
We also derive expressions for the relative oscillator strengths
of the excitonic transitions induced by circularly polarized
light. We show that the oscillator strengths are related to
derivatives of certain projections of the exciton wave function
at zero separation between electron and hole.

The article is organized as follows.
In Sec.~\ref{sec:theory}, we derive the theory for excitons in cuprous
oxide QWs taking into account the complex valence band structure. We
obtain the Hamiltonian, expressions for relative oscillator strengths
and provide a numerical algorithm for the computation of exciton
spectra and wavefunctions.
Results are presented and discussed in Sec.~\ref{sec:results}.
Concluding remarks are given in Sec.~\ref{sec:conclusion}.

\section{Theory and methods}
\label{sec:theory}
\begin{figure}
\includegraphics[width=0.98\columnwidth]{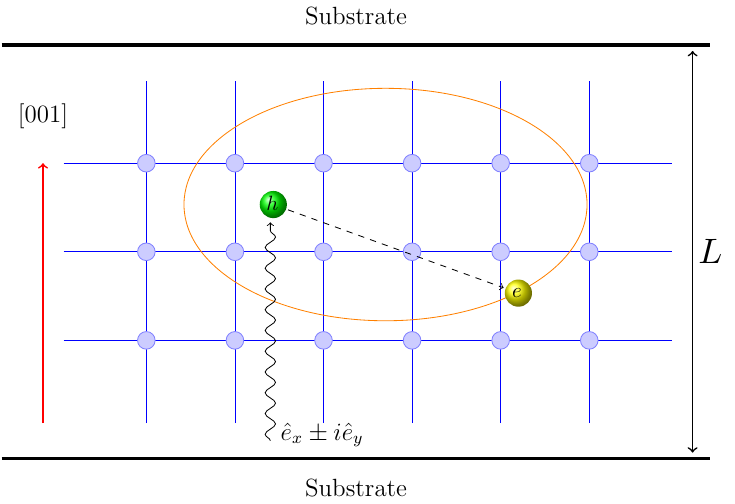}
 \caption{Sketch of the general setup for the creation of excitons in
   a cuprous oxide QW.  The $[001]$ axis of the crystal is aligned
   perpendicular to the QW plane.  Excitons are excited by circularly
   polarized light along this axis.}
\label{fig:crystal}
\end{figure}

In this section, we describe excitons in cuprous oxide QWs by taking into
account the complex structure of the valence band.
The general setup of our system is illustrated in Fig.~\ref{fig:crystal}.
Let $L$ be the width of a thin layer of cuprous oxide sandwiched between
another material with larger band gap. Due to band offsets at the
heterojunction, this system can be regarded as a QW.
Excitons in the QW are created by circularly polarized laser
light propagating perpendicularly to the QW plane.
In general, the exciton spectra depend on the orientation of the
crystal relative to the QW.
In this article, we consider a geometry in which the principal axis $[001]$
of the crystal is aligned along the $z$ direction perpendicular to the
QW plane.
The Hamiltonian including the impact of the valence band is
derived in Sec.~\ref{sec:Hamiltonian} and its symmetry properties are
discussed in Sec.~\ref{sec:symmetries}.
The computation of the oscillator strengths and the numerical algorithm
for the diagonalization of the Hamiltonian using an appropriate basis
set are explained in Secs.~\ref{sec:oscillator_strengths} and
\ref{sec:numerics}, respectively.

\subsection{Hamiltonian}
\label{sec:Hamiltonian}
The Hamiltonian for electron and hole in the QW is given by
\begin{align}
     H &= H_{\mathrm{e}}(\boldsymbol{p}_{\mathrm{e}})
       + H_{\mathrm{h}}(\boldsymbol{p}_{\mathrm{h}})
       - \frac{e^2}{4\pi\varepsilon_0\varepsilon}
         \frac{1}{|\boldsymbol{r}_{\mathrm{e}}-\boldsymbol{r}_{\mathrm{h}}|}\nonumber\\
       &+ V_{\mathrm{e}}(z_{\mathrm{e}}) + V_{\mathrm{h}}(z_{\mathrm{h}}),
\end{align}
where
\allowdisplaybreaks[2]
\begin{align}
     H_{\mathrm{e}}(\boldsymbol{p}_{\mathrm{e}}) &=
        E_{\mathrm{g}} + \frac{\boldsymbol{p}_{\mathrm{e}}^2}{2m_{\mathrm{e}}}\, , \\
     H_{\mathrm{h}}(\boldsymbol{p}_{\mathrm{h}}) &= H_{\mathrm{SO}}+\frac{1}{2\hbar^{2}m_0}\Big[ \hbar^{2}\left(\gamma_{1}+4\gamma_{2}\right)\boldsymbol{p}_{\mathrm{h}}^{2}\nonumber\\
     &+2\left(\eta_{1}+2\eta_{2}\right)\boldsymbol{p}_{\mathrm{h}}^{2}\left(\boldsymbol{I}\cdot\boldsymbol{S}_{\mathrm{h}}\right)
   -6\gamma_{2}\left(p_{1\mathrm{h}}^{2}{I}_{1}^{2}+\mathrm{c.p.}\right)\nonumber\\
   & -12\eta_{2}\left(p_{1\mathrm{h}}^{2}{I}_{1}{S}_{\mathrm{h}1}+\mathrm{c.p.}\right)\nonumber\\
   & -12\gamma_{3}\left[\left\{ p_{1\mathrm{h}},p_{2\mathrm{h}}\right\} \left\{ {I}_{1},{I}_{2}\right\} +\mathrm{c.p.}\right]\nonumber \\
   &  -12\eta_{3}\left[\left\{ p_{1{\mathrm{h}}},p_{2{\mathrm{h}}}\right\} \left({I}_{1}{S}_{\mathrm{h}2}+{I}_{2}{S}_{\mathrm{h}1}\right)+\mathrm{c.p.}\right]\Big]\, ,\\
     H_{\mathrm{SO}} &= \frac{2}{3}\Delta
        \left(1+\frac{1}{\hbar^2}\boldsymbol{I}\cdot\boldsymbol{S}_{\mathrm{h}}\right)\, , \\
\label{eqV}
  V_{\mathrm{e,h}}(z_{\mathrm{e,h}}) &= \left\{
  \begin{array}{lr}
    0  & \mbox{ if  } |z_{\mathrm{e,h}}| < L/2 \\
    \infty & \mbox{ if  } |z_{\mathrm{e,h}}| \ge L/2
  \end{array}
\right. .
\end{align}
Here, $E_{\mathrm{g}}$ is the gap energy, $m_0$ is the free electron mass,
and $\gamma_i, \eta_i$ are the Luttinger parameters. 
\begin{figure}
\includegraphics[width=0.9\columnwidth]{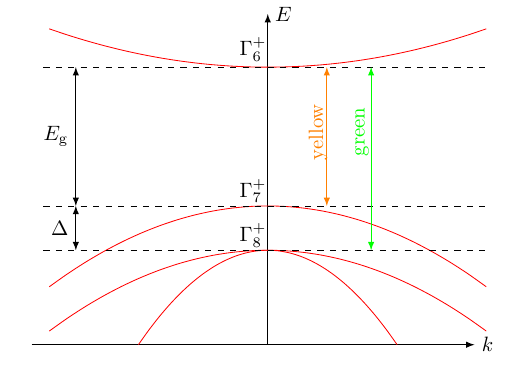}
\caption{Sketch of the band structure of the $\Gamma_6^+$ conduction bands, as well as
  the $\Gamma_7^+$ and $\Gamma_8^+$ valence bands near the $\Gamma$ point.}
\label{fig:BandStructure}
\end{figure}
The material parameters of cuprous oxide are given in
Table~\ref{tab:material_parameters}.
$\boldsymbol{I}$ and $\boldsymbol{S}_{\mathrm{h}}$ are the quasispin
with $I=1$ and the hole spin with $S_{\mathrm{h}}=1/2$, whose components
$I_i$ and $S_{\mathrm{h}i}$ with $i=1,2,3$
satisfy the commutation relations of an angular momentum or spin,
$\Delta$ is the spin-orbit coupling parameter,
$\{a,b\}=(ab+ba)/2$ is the symmetrized product, and c.p.\ denotes
cyclic permutation.
\begin{table}
\caption{Material parameters of Cu$_2$O used in the calculations.}
\vspace{1ex}
\begin{tabular}{l|lc}
  \hline
  Energy gap  & $E_{\rm g}=2.17208\,$eV & \cite{Kaz14}\\
  Spin-orbit coupling       & $\Delta=0.131\,$eV &\cite{Schoene16}\\
  Effective electron mass  & $m_{\rm e}=0.99m_0$ & \cite{Hodby_1976} \\
  Dielectric constant      & $\varepsilon=7.5$ &\cite{LandoltBornstein1998DielectricConstant}\\
  Valence band parameters & $\gamma_1=1.76$&\cite{Schoene16}\\
  ~~& $\gamma_2=0.7532$&\cite{Schoene16}\\
  ~~& $\gamma_3=-0.3668$&\cite{Schoene16}\\
  ~~& $\eta_1=-0.020$&\cite{Schoene16}\\
  ~~& $\eta_2=-0.0037$&\cite{Schoene16}\\
  ~~& $\eta_3=-0.0337$&\cite{Schoene16}\\
  \hline
\end{tabular}
\label{tab:material_parameters}
\end{table}

The potentials $V_{\mathrm{e},\mathrm{h}}(z_{\mathrm{e},\mathrm{h}})$
confine the motion of electron and hole across the QW along the $z$
direction.
For a finite-sized crystal surrounded by vacuum or air, the confinement
potential of the QW can be well approximated by an infinite potential
well, given in Eq.~\eqref{eqV}.
The energies of electron and hole states in such a QW,
\begin{equation}
  E_{N_{\mathrm{e}},N_{\mathrm{h}}}(L) = \frac{\hbar^2\pi^2}{2L^2}
  \left(\frac{N_{\mathrm{e}}^2}{m_{\mathrm{e}}}+\frac{N_{\mathrm{h}}^2}{m_{\mathrm{h}}}\right),
\label{eq:E_QW}
\end{equation}
can be described by two quantum numbers $N_{\mathrm{e}},N_{\mathrm{h}}=1,2,3,\dots$.
Here, $m_\mathrm{e} = 0.99m_0$ is the electron mass in the crystal and
$m_{\mathrm{h}} = m_0/\gamma_1$ is the hole mass.
The quantum-confinement energies~\eqref{eq:E_QW} play the role of scattering
thresholds, where bound states or resonances can accumulate~\cite{Scheuler2024}.

The inclusion of the full set of Luttinger parameters leads to the
non-trivial band structure shown in Fig.~\ref{fig:BandStructure}.
Using the notation for the symmetry group $O_\mathrm{h}$~\cite{Koster1963},
we can classify the bands as in the bulk crystal.
The lowest conduction band has the symmetry $\Gamma_6^+$, whereas the
uppermost valence bands are split into the uppermost $\Gamma_7^+$
band associated with the yellow exciton Rydberg series, and
the lower $\Gamma_8^+$ bands associated with the green series.
Note that for excitons in the QW we use the same values of the
Luttinger parameters as in the bulk.

We use the same dielectric constant $\varepsilon=7.5$ for
the crystal and the substrate and thus do not consider the dielectric
contrast described by the Rytova-Keldysh potential~\cite{Rytova,Keldysh}
for the interaction between electron and hole as discussed, e.g., in
Ref.~\cite{Belov2024}.
We also do not consider central-cell corrections, which can affect
even-parity excitons~\cite{Schweiner17b}.

In contrast to the bulk with its translational invariance in three
dimensions, for excitons in QWs there is a translational
invariance only in the two-dimensional $(x,y)$ QW plane.
We introduce relative and center-of-mass coordinates
\begin{subequations}
\label{eq:rel_coord}
\begin{align}
  x &= x_{\mathrm{e}}-x_{\mathrm{h}}\;,\;
  y = y_{\mathrm{e}}-y_{\mathrm{h}}\;,\\
  X &= \frac{m_{\mathrm{e}}x_{\mathrm{e}}+m_{\mathrm{h}}x_{\mathrm{h}}}{m_{\mathrm{e}}+m_{\mathrm{h}}}\;,\;
  Y = \frac{m_{\mathrm{e}}y_{\mathrm{e}}+m_{\mathrm{h}}y_{\mathrm{h}}}{m_{\mathrm{e}}+m_{\mathrm{h}}}
\end{align}
\end{subequations}
in this plane.
Using relative and center-of-mass momenta, $\bm{p}_{(2D)}=(p_1,p_2)$
and $\bm{P}_{(2D)}=(P_1,P_2)$, respectively, and additionally setting the
conserved center-of-mass momenta to zero, $P_1=P_2=0$, we obtain
\begin{widetext}
\allowdisplaybreaks[2]
\begin{align}
  &H(x,y,z_{\mathrm{e}},z_{\mathrm{h}},\bm{I},\bm{S}_\text{h})
    = E_\text{g}+H_\text{SO}-\frac{e^2}{4\pi\varepsilon_0\varepsilon}\frac{1}{\sqrt{x^2+y^2+(z_\text{e}-z_\text{h})^2}}+V_\text{e}(z_\text{e})+V_\text{h}(z_\text{h}) +\frac{p^2_\text{ez}}{2m_\text{e}}\nonumber\\
  &+\left[\frac{\gamma'_1}{2m_0}+\frac{2\gamma_2}{m_0}+\frac{\eta_1+2\eta_2}{\hbar^2m_0}(\bm{I}\cdot\bm{S_\text{h}})\right]\bm{p}_{(2D)}^2
  +\left[\frac{\gamma_1}{2m_0}+\frac{2\gamma_2}{m_0}+\frac{\eta_1+2\eta_2}{\hbar^2m_0}(\bm{I}\cdot\bm{S_\text{h}})-\frac{3\gamma_2}{\hbar^2m_0}I_3^2-\frac{6\eta_2}{\hbar^2m_0}I_3S_\text{h3}\right]p_\text{hz}^2\nonumber\\
  &-\frac{3\gamma_2}{\hbar^2m_0}\left(p_1^2I_1^2+p_2^2I_2^2\right)-\frac{6\gamma_3}{\hbar^2m_0}\left[\{p_1,p_2\}\{I_1,I_2\}\right]
  +\frac{6\gamma_3}{\hbar^2m_0}\left[\{p_2,p_\text{hz}\}\{I_2,I_3\}+ \{p_\text{hz},p_1\}\{I_3,I_1\}\right]\nonumber\\
  &-\frac{6\eta_2}{\hbar^2m_0}(p_1^2I_1S_\text{h1}+p_2^2I_2S_\text{h2})
  -\frac{6\eta_3}{\hbar^2m_0}\left[\{p_1,p_2\}(I_1S_\text{h2}+I_2S_\text{h1})\right]\nonumber\\
  &+\frac{6\eta_3}{\hbar^2m_0}\left[\{p_2,p_\text{hz}\}(I_2S_\text{h3}+I_3S_\text{h2})+\{p_\text{hz},p_1\}(I_3S_\text{h1}+I_1S_\text{h3})\right]
\label{eq:H_cartesian}
\end{align}
with $\gamma'_1=\gamma_1+m_0/m_{\mathrm{e}}$.
When comparing the Hamiltonian~\eqref{eq:H_cartesian} with the
two-band model for excitons in a QW
\cite{Scheuler2024,Belov2024,Kuehner2025} it is important to note that
not only the dimension of the Hamiltonian is increased by the
additional spin degrees of freedom, i.e., the quasispin $\bm{I}$ and
hole spin $\bm{S}_{\mathrm{h}}$ describing the valence band, but also
the coordinate space is increased from three degrees of freedom
$(\rho,z_{\mathrm{e}},z_{\mathrm{h}})$ in the two-band model to four
degrees of freedom $(x,y,z_{\mathrm{e}},z_{\mathrm{h}})$ when
considering the complete structure of the valence band.
This is a consequence of the reduced symmetry of the
Hamiltonian~\eqref{eq:H_cartesian} as discussed in
Sec.~\ref{sec:symmetries}.

The Hamiltonian in the form of Eq.~\eqref{eq:H_cartesian} with
Cartesian coordinates is not well suited for the computation of matrix
elements in the basis introduced in Sec.~\ref{sec:numerics}, which
takes advantage of the polar coordinates $(\rho,\phi)$ in the QW plane
and contains angular momentum functions $\exp(im\phi)$ and spin states
$|J,m_J\rangle$ for the coupled quasispin and hole spin
$\boldsymbol{J}=\boldsymbol{I}+\boldsymbol{S}_{\mathrm{h}}$.
For this basis, it is more convenient to introduce creation and annihilation
operators
\begin{align}
    I_\pm &= I_1 \pm iI_2\, ,\quad
    S_{\mathrm{h}\pm} = S_{\mathrm{h}1} \pm iS_{\mathrm{h}2}\, ,\quad
    p_\pm = p_1 \pm ip_2\, ,
\end{align}
where the operators $I_\pm$ and $S_{\mathrm{h}\pm}$ couple states with
$\Delta m_J=\pm 1$ and the operators $p_\pm$ couple states with
$\Delta m=\pm 1$.
Representing the Hamiltonian~\eqref{eq:H_cartesian} in terms of
these operators and sorting terms with respect to their action on the
angular momentum and spin part of the basis states, the complete
Hamiltonian can finally be written as
\begin{equation}
    H = H_{\mathrm{diag}} + H_{\Delta m=0} + H_{\Delta m=1} + H_{\Delta m=2} + H_{\Delta m_F=4},
\label{eq:H_final}
\end{equation}
where
\begin{subequations}
\allowdisplaybreaks[2]
  \begin{align}
    \label{eq:H_diag}
       &H_{\mathrm{diag}} = E_{\mathrm{g}} + H_{\mathrm{SO}} + V_{\mathrm{C}}
      + \frac{1}{2m_{\mathrm{e}}}p_{\mathrm{e}z}^2
      + \frac{\gamma_1'}{2m_0}p_+p_- + \frac{\gamma_1}{2m_0}p_{\mathrm{h}z}^2
      + \left[\frac{2\gamma_2}{m_0}+\frac{\eta_1+2\eta_2}{\hbar^2m_0}
      (\boldsymbol{I}\cdot\boldsymbol{S}_{\mathrm{h}}) \right]
      \left(p_+p_-+p_{\mathrm{h}z}^2\right)\, ,\\
\label{eq:H_Dm=0}
    &H_{\Delta m=0} = \left[-\frac{3\gamma_2}{2\hbar^2m_0}\{I_+,I_-\}
      -\frac{3\eta_2}{2\hbar^2m_0}(I_+S_{\mathrm{h}-}+I_-S_{\mathrm{h}+})\right]p_+p_-
      + \left[-\frac{3\gamma_2}{\hbar^2m_0}I_3^2
      -\frac{6\eta_2}{\hbar^2m_0}I_3S_{\mathrm{h}3}\right]p_{\mathrm{h}z}^2\, ,\\
    &H_{\Delta m=1} = \frac{3\gamma_3}{\hbar^2m_0}\left(\{I_3,I_+\}p_{\mathrm{h}z}p_-+\mathrm{h.c.}\right)
      + \frac{3\eta_3}{\hbar^2m_0}\left[\left(I_3S_{\mathrm{h}+}+I_+S_{\mathrm{h}3}\right)p_{\mathrm{h}z}p_-
         +\mathrm{h.c.}\right]\, ,\\
    &H_{\Delta m=2} = -\frac{3(\gamma_2+\gamma_3)}{8\hbar^2m_0}\left(I_+^2p_-^2+\mathrm{h.c.}\right)
      -\frac{3(\eta_2+\eta_3)}{4\hbar^2m_0}\left(I_+S_{\mathrm{h}+}p_-^2+\mathrm{h.c.}\right) \, ,\\
    &H_{\Delta m_F=4} = -\frac{3(\gamma_2-\gamma_3)}{8\hbar^2m_0}\left(I_-^2p_-^2+\mathrm{h.c.}\right)
      -\frac{3(\eta_2-\eta_3)}{4\hbar^2m_0}\left(I_-S_{\mathrm{h}-}p_-^2+\mathrm{h.c.}\right) \, .
\end{align}
\end{subequations}
\end{widetext}
Here, $V_{\mathrm{C}}$ is the Coulomb potential and h.c.\ denotes the
Hermitian conjugate.
The operator $H_{\mathrm{diag}}$ is diagonal with respect to the
angular momentum and spin quantum numbers $m$, $J$, and
$m_J=m_I+m_{S_{\mathrm{h}}}$.
For an appropriately chosen effective hole mass this part of the
Hamiltonian describes the exciton within a hydrogenlike two-band
model as discussed in Refs.~\cite{Scheuler2024,Belov2024,Kuehner2025}.
The operator $H_{\Delta m=0}$ couples basis states with quantum
numbers $J=1/2$ and $J=3/2$ but the same angular momentum quantum
number $m$, i.e., this term causes a coupling between the yellow and
green exciton series without breaking the continuous rotational
symmetry around the $z$ axis.
The operators $H_{\Delta m=1}$ and $H_{\Delta m=2}$ couple states with
different angular momentum quantum numbers $m$, and thus $m$ is not a
good quantum number when considering the impact of the valence band.
Finally, the operator $H_{\Delta m_F=4}$ couples states with
$\Delta m_F=\pm 4$, where $m_F=m+m_J$ is the quantum number for the
$z$ component of the coupled spin and angular momenta
$\boldsymbol{F}=\boldsymbol{L}+\boldsymbol{J}
=\boldsymbol{L}+\boldsymbol{I}+\boldsymbol{S}_{\mathrm{h}}$.
Note that when using basis states with given quantum numbers $m_F$ the
matrix representation of the Hamiltonian~\eqref{eq:H_final} has a
block structure with four blocks, which can be diagonalized separately.
This block structure is directly related to the symmetry properties of the
Hamiltonian~\eqref{eq:H_final} as shown below in Sec.~\ref{sec:symmetries}.
Note also that $H_{\Delta m_F=4}$ would vanish in case of
equal Luttinger parameters $\gamma_2=\gamma_3$ and $\eta_2=\eta_3$,
i.e., $m_F$ would be an exact quantum number in that case.

\subsection{Symmetry properties}
\label{sec:symmetries}
\begin{figure}
  \includegraphics[width=0.9\columnwidth]{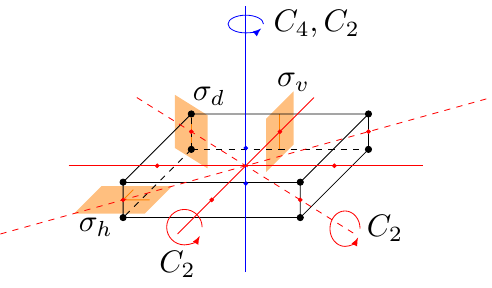}
  \caption{Illustration of symmetry operations of the group
    $D_{4\mathrm{h}}$. $C_2$ and $C_4$ denote two- and fourfold
    rotations, respectively, $\sigma_h$, $\sigma_v$, and $\sigma_d$
    reflections with respect to different kinds of symmetry planes. In
    addition to the depicted operations, there is also the inversion
    $I$ and the improper rotation $S_4 = \sigma_h C_4 $.}
\label{fig:symmetries}
\end{figure}

In comparison to the hydrogenlike model of excitons in a
QW~\cite{Kuehner2025}, the inclusion of the complex 
valence band leads to a further reduction of the symmetry.
We consider the situation in which the [001] axis of the crystal is
perpendicular to the QW plane.
In this case, the symmetry of the system is reduced from $D_{\infty\mathrm{h}}$,
and the complete Hamiltonian~\eqref{eq:H_final} is now only invariant
under the symmetry operations of the $D_{4\mathrm{h}}$ symmetry
group~\cite{Koster1963}, containing rotations $C_4$ and $C_2$, inversion
$I$, improper rotations $S_4$, and reflections $\sigma_{\mathrm{h}}$,
$\sigma_{\mathrm{v}}$, and $\sigma_{\mathrm{d}}$,
as illustrated in Fig.~\ref{fig:symmetries}.
With $m_J=m_I+m_{S_{\mathrm{h}}}$ and $m_F=m+m_J$, the $C_4$ symmetry
implies that the Hamiltonian only couples basis states with $\Delta
m_F=0,\pm 4$.
Note that this symmetry-derived constraint on matrix elements
only applies to the total angular momentum, i.e.,
individual angular momenta like $m$ or $m_J$ are not restricted in
this way in their possible couplings, as can be seen by the terms in
the Hamiltonian~\eqref{eq:H_final}.
Due to the reduced symmetry of the system,
the eigenstates can no longer be classified according to their angular
momentum quantum numbers, but instead by the irreducible
representations of the symmetry group
$D_{4\mathrm{h}}$, or rather, as there are also spin degrees of
freedom, the corresponding double group to $D_{4\mathrm{h}}$.
Without consideration of the electron spin $\boldsymbol{S}_{\mathrm{e}}$,
the total angular momentum $\boldsymbol{F}$ in the system is
half-integer valued, and only the irreducible representations
$\Gamma_6^+$, $\Gamma_6^-$, $\Gamma_7^+$, and $\Gamma_7^-$ are possible for the
eigenstates~\cite{Koster1963}, where the superscript $+$ or $-$
indicates the total parity of the corresponding eigenstates under
inversion $I$.
The eigenstates are superpositions of states with $m_F$ differing by
multiples of 4, and thus can be assigned to one of the following groups
\begin{subequations}
\label{eq:m_F-all}
\begin{align}
\label{eq:m_F-Gamma_6}
  m_F &= \left\{
        \begin{array}{l}
          \dots,-\frac{13}{2},-\frac{5}{2},\frac{3}{2},\frac{11}{2},\dots\\[1ex]
          \dots,-\frac{11}{2},-\frac{3}{2},\frac{5}{2},\frac{13}{2},\dots
        \end{array} \right.\\
\label{eq:m_F-Gamma_7}
  m_F &= \left\{
        \begin{array}{l}
          \dots,-\frac{15}{2},-\frac{7}{2},\frac{1}{2},\frac{9}{2},\dots\\[1ex]
          \dots,-\frac{9}{2},-\frac{1}{2},\frac{7}{2},\frac{15}{2},\dots
        \end{array} \right.
\end{align}
\end{subequations}
Here, blocks related by a sign change of $m_F$ are degenerate due to
time-reversal symmetry.
Equations~\eqref{eq:m_F-Gamma_6} and \eqref{eq:m_F-Gamma_7}
describe twofold degenerate states in subspaces $\Gamma_6^\pm$ and
$\Gamma_7^\pm$, respectively, as can be seen as follows.

To understand how these blocks relate to the irreducible
representations $\Gamma_{6,7}^\pm$, we can look at a state with total
angular momentum $F=1/2$.
We need to take into account that the eigenstates of the quasispin $I$ 
do not behave exactly like eigenstates of a standard angular
momentum. In $D_{4\mathrm{h}}$, the eigenstates of the quasispin
transform like the sum of irreducible representations $\Gamma_4^+ +
\Gamma_5^+ = (\Gamma_2^+ + \Gamma_5^+)\times\Gamma_3^+$, while
the eigenstates of a standard spin of 1 behave like $\Gamma_2^+ +
\Gamma_5^+$. Hence, one obtains the  total symmetry by the usual addition 
of angular momenta, followed by a multiplication with $\Gamma_3^+$. 
With this prescription in mind, we can use that an angular momentum of 
$1/2$ transforms according to $\Gamma_6^\pm$ in $D_{4\mathrm{h}}$~\cite{Koster1963}.
Multiplying by $\Gamma_3^+$, we find that a state with total angular
momentum $F=1/2$, and thus states with $m_F = 1/2$ and $m_F = -1/2$
transform according to $\Gamma_7^\pm$.

For the computation of oscillator strengths in Sec.~\ref{sec:oscillator_strengths},
the electron spin $\boldsymbol{S}_{\mathrm{e}}$ must also be taken into account.
The total symmetries of the system including the electron spin are
discussed below in Sec.~\ref{sec:selection_rules}.

\subsection{Oscillator strengths}
\label{sec:oscillator_strengths}
For the computation of oscillator strengths we generalize the theory
developed in Ref.~\cite{Kuehner2025} within a two-band model.
We consider excitons in the many-particle framework.
The ground state $|\Phi_0 \rangle$ of the crystal is the state where
all electrons of the crystal are in the valence bands, while all
conduction bands are empty.
An exciton is created when one of these electrons is lifted from the
valence bands into a conduction band.
The resulting state can be written as
\begin{equation}
  |\psi^{\sigma \tau}_{vc,\nu} \rangle = \sum_{\boldsymbol{q}_c,\boldsymbol{q}_v} f^{\tau,\sigma}_{vc,\nu}(\boldsymbol{q}_c,\boldsymbol{q}_v) c^\dagger_{c,\boldsymbol{q}_c,\tau} c_{v,\boldsymbol{q}_v,\sigma} | \Phi_0 \rangle\,.
\label{eq:ExcitonManyBody}
\end{equation}
Here, $c^\dagger_{c,\boldsymbol{q}_c}$ denotes the creation operator for an electron
with crystal momentum $\boldsymbol{q}_c$ and electron spin $\tau$ in
the conduction band $c$, and $c_{v,\boldsymbol{q}_v}$ is the analogous
annihilation operator in the valence band $v$, which creates a hole
with crystal momentum $-\boldsymbol{q}_v$ and hole spin $-\sigma$. The
coefficient $f^{\tau,\sigma}_{vc,\nu}(\boldsymbol{q}_c,\boldsymbol{q}_v)$ in the 
superposition is the envelope function, which is obtained as the solution of the
exciton Schrödinger equation with Hamiltonian~\eqref{eq:H_final}
transformed into crystal-momentum space.
When considering the complex valence band structure, $c$ and $v$ are
not good quantum numbers, and the correct states are obtained by
forming superpositions. The subscript $\nu$ denotes further quantum
numbers or indices associated with the state.

Using relative and center-of-mass coordinates in the QW plane [see
Eqs.~\eqref{eq:rel_coord}], we can express
the $x$ and $y$ components of $\boldsymbol{q}_c$ and
$\boldsymbol{q}_v$ through the center-of-mass momentum
$\boldsymbol{K}_\mathrm{2D}$ and the relative in-plane
momentum $\boldsymbol{q}$.
As $\boldsymbol{K}_\mathrm{2D}$ is conserved, we are discussing
states $\ket{\psi_{\nu \boldsymbol{K}_\mathrm{2D}}}$, where the sum in
Eq.~\eqref{eq:ExcitonManyBody} is restricted to a fixed value of
$\boldsymbol{K}_\mathrm{2D}$.
Combining the symmetry considerations and general selection rules in
Refs.~\cite{Kuehner2025,Koster1963}, the transition matrix elements
fulfill
\begin{subequations}
\label{eq:TransitionRulesPolarization}
\begin{align}
  M_x &\propto (\partial_y\bra{u_{xy\boldsymbol{0}}} +
      \partial_z\bra{u_{zx\boldsymbol{0}}})\bra{S=0}\ket{\psi_{\nu \boldsymbol{K}_\mathrm{2D}}}\bigr\rvert_{\boldsymbol{r} = 0},\\
  M_y &\propto (\partial_z\bra{u_{yz\boldsymbol{0}}} +
      \partial_x\bra{u_{xy\boldsymbol{0}}})\bra{S=0}\ket{\psi_{\nu \boldsymbol{K}_\mathrm{2D}}}\bigr\rvert_{\boldsymbol{r} = 0},\\
  M_z &\propto (\partial_x\bra{u_{zx\boldsymbol{0}}} +
      \partial_y\bra{u_{yz\boldsymbol{0}}})\bra{S=0}\ket{\psi_{\nu \boldsymbol{K}_\mathrm{2D}}}\bigr\rvert_{\boldsymbol{r} = 0}.
\end{align}
\end{subequations}
Here, the application of $\bra{S=0} = \bra{S=0,M_S=0}$ enforces the
absence of spin flips of the total spin
$\boldsymbol{S} = \boldsymbol{S}_\mathrm{e} + \boldsymbol{S}_\mathrm{h}$
with the electron spin $\boldsymbol{S}_\mathrm{e}$ added to the description.
The states $\ket{u_{i\boldsymbol{0}}}$ with $i$ = $yz$, $xz$, $xy$
denote the one-particle hole states in the valence band $i$ at the
$\Gamma$ point.
In the case of Cu$_2$O, these transform like the irreducible
representation $\Gamma_5^+$ of the cubic group $O_\mathrm{h}$. They
are related to the eigenstates of the quasispin $\boldsymbol{I}$
introduced in Sec.~\ref{sec:Hamiltonian} via~\cite{Schweiner17c}
\begin{subequations}
\label{eq:BlochFunctionQuasispin}
\begin{align}
&\ket{I=1,m_I = +1} = -(\ket{u_{yz\boldsymbol{0}}} + i \ket{u_{zx\boldsymbol{0}}})/\sqrt{2},\\
&\ket{I=1,m_I = 0} = \ket{u_{xy\boldsymbol{0}}},\\
&\ket{I=1,m_I = -1} = +(\ket{u_{yz\boldsymbol{0}}} - i \ket{u_{zx\boldsymbol{0}}})/\sqrt{2}.
\end{align}
\end{subequations}
We are specifically interested in the oscillator strengths for
circularly polarized light.
We thus need to take suitable linear combinations of the transition
matrix elements in Eqs.~\eqref{eq:TransitionRulesPolarization}
\begin{align}
M_{\pm} &\propto ( (i \partial_y \mp  \partial_x) \bra{m_I = 0,S=0}\nonumber\\ 
        & +\sqrt{2} \partial_z \bra{m_I = \pm 1,S=0} ) \ket{\psi_{\nu \boldsymbol{K}_\mathrm{2D}}}\bigr\rvert_{\boldsymbol{r} = 0}\,.
\label{eq:TransitionsmatrixelementCircular}
\end{align}
Other polarizations can be evaluated analogously.

Using these relations and the Clebsch-Gordan coefficients, the spin
and quasispin bra vectors can be rewritten as
\begin{equation}
\ket{S=0,m_I} = \sum_{J,m_J,m_{S_\mathrm{e}}} c^{m_I}_{J,m_J,m_{S_\mathrm{e}}} \ket{J,m_J,m_{S_\mathrm{e}}},
\end{equation}
with the coefficients
\begin{equation}
  c^{m_I}_{J,m_J,m_{S_\mathrm{e}}}=\bra{J,m_J,m_{S_\mathrm{e}}}\ket{S=0, M_S=0; I, m_I}  \, .
  \label{eq:BasisChangeCoefficients}
\end{equation}
Generalizing the result from Ref.~\cite{Kuehner2025} when $m$ is not a
good quantum number, the spatial part of the wave function
$\ket{\psi^{\sigma\tau}_{vc,\nu \boldsymbol{K}_\mathrm{2D}}}$ can be written as
\begin{align}
&\bra{J,m_J,m_{S_\mathrm{e}}}\bra{\boldsymbol{K},\rho,\phi,z}\ket{\psi_{\nu \boldsymbol{K}_\mathrm{2D}}} \nonumber\\
= \;&\sum_m \int_{-\frac{L}{2}}^\frac{L}{2} \psi_{m,J,m_J,m_{S_\mathrm{e}}}^{\nu \boldsymbol{K}_\mathrm{2D}}(\rho,z,Z) \frac{e^{im\phi}}{\sqrt{2\pi L}}e^{-iK_z Z} \mathrm{d}Z.
\label{eq:ProjectedCoordinateWavefunction}
\end{align}
The photon momenta for dipole transitions to yellow excitons
($\lambda\approx 580\,$nm) are negligibly small. We can thus set $K_z$ and
$\boldsymbol{K}_\mathrm{2D}$ to zero and ignore it in the following.
Common prefactors can also be dropped in the calculation of relative oscillator strengths.
From the condition of continuity at the origin, we can deduce for the
derivatives that~\cite{Kuehner2025}
\begin{subequations}
\label{eq:DerivativesOsc}
\begin{align}
&\frac{\partial}{\partial x} \left(\psi_{m,J,m_J,m_{S_\mathrm{e}}}^\nu(\rho,z,Z)e^{i m \phi}\right)\Bigr\rvert_{\rho=z=0}\nonumber \\
= \;&\delta_{\abs{m},1}\partial_\rho \psi_{m,J,m_J,m_{S_\mathrm{e}}}^\nu(\rho,z,Z)\bigr\rvert_{\rho=z=0}\,,\\[1ex]
&\frac{\partial}{\partial y} \left(\psi_{m,J,m_J,m_{S_\mathrm{e}}}^\nu(\rho,z,Z)e^{i m \phi}\right)\Bigr\rvert_{\rho=z=0}\nonumber \\
 = \;&i m \delta_{\abs{m},1} \partial_\rho \psi_{m,J,m_J,m_{S_\mathrm{e}}}^\nu(\rho,z,Z)\bigr\rvert_{\rho=z=0}\,,\\[1ex]
&\frac{\partial}{\partial z} \left(\psi_{m,J,m_J,m_{S_\mathrm{e}}}^\nu(\rho,z,Z)e^{i m \phi}\right)\Bigr\rvert_{\rho=z=0}\nonumber \\
 = \;&\delta_{m,0} \partial_z \psi_{m,J,m_J,m_{S_\mathrm{e}}}^\nu(\rho,z,Z)\bigr\rvert_{\rho=z=0}\,.
\end{align}
\end{subequations}
To obtain the oscillator strengths for the excitation of the
twofold generate states in the irreducible subspaces with circularly
polarized light, the contributions of both states must be added.
Combining Eqs.~\eqref{eq:TransitionsmatrixelementCircular},
\eqref{eq:ProjectedCoordinateWavefunction}, and \eqref{eq:DerivativesOsc},
and using the time-reversal symmetry between the two degenerate
states discussed in Sec.~\ref{sec:symmetries},
the desired oscillator strengths can be evaluated as
\begin{align}
f_{\mathrm{rel}} &= \bigg{|}\int_{-\frac{L}{2}}^\frac{L}{2}\left[ \mathcal{M}_\rho^+(Z) + \mathcal{M}_z^+(Z)\right] \mathrm{d}Z \bigg{|}^2\nonumber\\
  &+ \bigg{|}\int_{-\frac{L}{2}}^\frac{L}{2}\left[ \mathcal{M}_\rho^-(Z) + \mathcal{M}_z^-(Z)\right] \mathrm{d}Z \bigg{|}^2
\label{eq:relativeOscillatorStrength_integral}
\end{align}
with the transition amplitude densities
\begin{subequations}
\label{eq:transition_amplitudes}
\begin{align}
  \label{eq:transition_amplitudes1}
\mathcal{M}^{\pm}_{\rho}(Z) &= \sqrt{2} \sum_{J,m_J} c^{m_I = 0,\ast}_{J,m_J,m_{S_\mathrm{e}}}\partial_\rho \psi_{\pm 1,J,m_J,m_{S_\mathrm{e}}}^\nu(\rho,z,Z)\Bigr\rvert_{\rho=z=0}\,, \\\label{eq:transition_amplitudes2}
\mathcal{M}^{\pm}_{z}(Z) &= \sum_{J,m_J} c^{m_I = \pm 1,\ast}_{J,m_J,m_{S_\mathrm{e}}} \partial_z \psi_{0,J,m_J,m_{S_\mathrm{e}}}^\nu(\rho,z,Z)\Bigr\rvert_{\rho=z=0}\,,
\end{align}
\end{subequations}
and the complex conjugate of the coefficients given in
Eq.~\eqref{eq:BasisChangeCoefficients}.
Note that the oscillator strengths $f_{\mathrm{rel}}$ in
Eq.~\eqref{eq:relativeOscillatorStrength_integral} do not depend on
the handedness of the circularly polarized light.
Details on the numerical evaluation of
Eqs.~\eqref{eq:relativeOscillatorStrength_integral} and
\eqref{eq:transition_amplitudes} are presented in
Sec.~\ref{sec:numerics}.

\subsubsection{Selection rules for dipole transitions}
\label{sec:selection_rules}
As the dipole operator has odd parity, and parity is a good quantum
number in the system, we can deduce that only states with odd total
parity belonging to the irreducible representations $\Gamma_6^-$ and
$\Gamma_7^-$ can be optically excited in the setup 
shown in Fig.~\ref{fig:crystal}. Correspondingly,
the integrals in Eq.~\eqref{eq:relativeOscillatorStrength_integral}
must vanish for states with even total parity.

Precise selection rules can be deduced by application of group
theory~\cite{Koster1963} when considering the symmetry of the dipole
operator and the respective symmetries of the states.
In the context of oscillator strengths, we add the
electron spin symmetry $\Gamma_6^+$, leading to the total symmetries
that decompose into a sum of subspaces as
\begin{subequations}
\label{eq:StatesTotalDegenerateSpaces}
\begin{align}
  \Gamma_6^\pm \times \Gamma_6^+ &= \Gamma_1^\pm + \Gamma_2^\pm +\Gamma_5^\pm\,,\\
  \Gamma_7^\pm \times \Gamma_6^+ &= \Gamma_3^\pm + \Gamma_4^\pm + \Gamma_5^\pm\,.
\end{align}
\end{subequations}
The dipole operator $D^-_{l = 1}$ on the other hand decomposes
into the sum $\Gamma_2^- + \Gamma_5^-$ of irreducible representations.
Here, $\Gamma_2^-$ corresponds to linear polarization along the $z$ direction,
whereas the two-dimensional $\Gamma_5^-$ corresponds to polarizations
in the QW plane, including the circular polarizations we
consider here.
From this, it follows that light circularly polarized in the $xy$ plane can
create excitons in the twofold degenerate $\Gamma_5^-$ spaces. As
$m_{S_\mathrm{e}}$ is a good quantum number, the entire
negative-parity subspaces in
Eqs.~\eqref{eq:StatesTotalDegenerateSpaces} are degenerate. As they
both fully contain $\Gamma_5^-$ spaces, it follows that they can both
be excited by left- and right-handed circularly polarized light.
In fact, from time-reversal symmetry of the system, it follows
that both polarizations lead to identical spectra.

Note that a more restricted selection rule is valid for $z$-polarized light. 
In that case, only $\Gamma_6^-$ states can be excited.
However, this would require light oriented in the QW plane, and thus a
different experimental setup than shown in Fig.~\ref{fig:crystal}.

\subsection{Numerical algorithm}
\label{sec:numerics}
To compute the exciton eigenenergies and eigenfunctions of the
Hamiltonian~\eqref{eq:H_final}, we extend the basis used in
the hydrogenlike model~\cite{Scheuler2024,Belov2024} for the wave
functions in the three-dimensional coordinate space
$(\rho,z_{\mathrm e}, z_{\mathrm h})$ by the additional
angular momentum and spin degrees of freedom.
We use the ansatz
\begin{equation}
  \ket{\Psi}
  = \sum_{J,m_J,m} \psi_{m,J,m_J}(\rho, z_{\mathrm e}, z_{\mathrm h})\,\ket{m,J,m_J} 
\label{eq:psi_total_expand}
\end{equation}
with the basis states
\begin{equation}
  |m,J,m_J\rangle=e^{im\phi}\,|J,m_J\rangle
  \label{eq:basis_states}
\end{equation}
as the product of the eigenstates of the angular momentum operator
$L_z$ and the eigenstates of the coupled quasispin and hole spin
$\bm{J}=\bm{I}+\bm{S}_{\mathrm{h}}$.
The wave functions $\psi_{m,J,m_J}(\rho, z_{\mathrm e}, z_{\mathrm h})$
in Eq.~\eqref{eq:psi_total_expand} are expanded over a basis of
B-splines of order $n$ as
\begin{align}
  &\psi_{m,J,m_J}(\rho, z_{\mathrm e}, z_{\mathrm h}) \nonumber\\
  = &\frac{1}{\sqrt{\rho}} \sum_{i,j,k} c^{m,J,m_J}_{ijk}\,
      B_i^{n}(\rho)\,B_j^{n}(z_{\mathrm e})\,B_k^{n}(z_{\mathrm h}) 
\label{eq:psi_coord_expand}
\end{align}
similar as in Refs.~\cite{Scheuler2024,Belov2024}.
For convenience, the definition and some properties
of B-splines are collected in Appendix~\ref{app:B-splines}.
Note that we do not expand the $\phi$ dependence in Eq.~\eqref{eq:psi_total_expand}
using B-splines but rather use the angular momentum states $\exp(im\phi)$
as basis states.

The expansions~\eqref{eq:psi_total_expand} and
\eqref{eq:psi_coord_expand} lead to a generalized eigenvalue problem
\begin{equation}
  \sum_\beta H_{\alpha\beta} c_\beta = E \sum_\beta O_{\alpha\beta} c_\beta
\label{eq:generalized_ev_problem}
\end{equation}
for the energy eigenvalues of the Hamiltonian~\eqref{eq:H_final}
and the expansion coefficients $c^{m,J,m_J}_{ijk}$ of the
corresponding eigenfunctions.
Here, the indices $\alpha$ and $\beta$ count both the
elements of the total basis set composed of B-splines, angular
momentum, and spin states in Eq.~\eqref{eq:psi_total_expand}
and the expansion coefficients $c^{m,J,m_J}_{ijk}$ when arranged as
one-dimensional vectors.

The elements of the matrices in Eq.~\eqref{eq:generalized_ev_problem}
are computed as follows.
The matrix elements of all terms of the Hamiltonian~\eqref{eq:H_final},
except for the Coulomb potential $V_C$, can be factorized into products of
one-dimensional integrals of B-splines or their derivatives along
coordinates $\rho$, $z_{\mathrm{e}}$, and $z_{\mathrm{h}}$, as well as
analytically pre-calculated matrix elements of spin operators.
Details of the computation of matrix elements of the momentum
operators and the spin operators are given in
Appendices~\ref{app:momentum_operators} and \ref{app:spin_operators},
respectively.
The matrix elements of the Coulomb potential $V_C$ are diagonal with
respect to the angular momentum and spin basis.
The non-trivial three-dimensional integrals in the
$(\rho,z_{\mathrm{e}},z_{\mathrm{h}})$ coordinate space are computed,
using the B-spline basis, as explained in Ref.~\cite{Scheuler2024}.

\begin{table}[b]
\caption{Selected angular momentum and spin basis states
  $|m,J,m_J\rangle$ with quantum numbers $m_F=-7/2$, $1/2$, and $9/2$
  belonging to the irreducible subspaces $\Gamma_7^\pm$ and used for
  numerical calculations.}
\vspace{1ex}
\begin{tabular}{c|c|c}
  $m_F=-\frac{7}{2}$ & $m_F=\frac{1}{2}$ & $m_F=\frac{9}{2}$ \\[0.5ex]
  \hline
  &&\\[-2ex]
  $|{-5},\frac{3}{2},+\frac{3}{2}\rangle$ & $|{-1},\frac{3}{2},+\frac{3}{2}\rangle$ & $|3,\frac{3}{2},+\frac{3}{2}\rangle$\\[0.5ex]
  $|{-4},\frac{3}{2},+\frac{1}{2}\rangle$ & $|{+0},\frac{3}{2},+\frac{1}{2}\rangle$ & $|4,\frac{3}{2},+\frac{1}{2}\rangle$\\[0.5ex]
  $|{-4},\frac{1}{2},+\frac{1}{2}\rangle$ & $|{+0},\frac{1}{2},+\frac{1}{2}\rangle$ & $|4,\frac{1}{2},+\frac{1}{2}\rangle$\\[0.5ex]
  $|{-3},\frac{1}{2},-\frac{1}{2}\rangle$ & $|{+1},\frac{1}{2},-\frac{1}{2}\rangle$ & $|5,\frac{1}{2},-\frac{1}{2}\rangle$\\[0.5ex]
  $|{-3},\frac{3}{2},-\frac{1}{2}\rangle$ & $|{+1},\frac{3}{2},-\frac{1}{2}\rangle$ & $|5,\frac{3}{2},-\frac{1}{2}\rangle$\\[0.5ex]
  $|{-2},\frac{3}{2},-\frac{3}{2}\rangle$ & $|{+2},\frac{3}{2},-\frac{3}{2}\rangle$ & $|6,\frac{3}{2},-\frac{3}{2}\rangle$
\end{tabular}
\label{tab:spin_basis}
\end{table}
The generalized eigenvalue problem~\eqref{eq:generalized_ev_problem}
is solved using ARPACK routines optimized for band matrices~\cite{arpackuserguide}.
The band and sparse structure of the resulting Hamiltonian matrix
$H_{\alpha\beta}$ and overlap matrix $O_{\alpha\beta}$ arises
naturally from the compact support of the B-spline functions.
We use 15 B-spline functions of order $n=5$ for both the $z_{\mathrm e}$
and $z_{\mathrm h}$ directions, defined on equidistant grids, and 26
B-spline functions for the $\rho$ direction.
For the $\rho$ coordinate, the grid nodes are distributed
non-equidistantly with a decreasing density over the interval
$[0,200\,\mathrm{nm}]$, providing a finer resolution near the origin
where both the potential and the wave function vary most rapidly, see 
Appendix~\ref{app:B-splines} for more details.
Because only basis states with $\Delta m_F=0,\pm4$ are coupled,
it is sufficient to use one of the sets in Eq.~\eqref{eq:m_F-all}
for the numerical diagonalization of the Hamiltonian.
  
We are mainly interested in the optically active states with
dominant $m=0$ and $m=1$ contributions.
Convergence of these states is achieved when restricting the spin
and angular momentum basis to the subset with $m_F=-7/2$, $1/2$, and
$9/2$ given in Table~\ref{tab:spin_basis} in the irreducible
subspace $\Gamma_7^\pm$ and to the subset with $m_F=-5/2$, $3/2$, and
$11/2$ in the subspace $\Gamma_6^\pm$.
States in the remaining subsets in Eq.~\eqref{eq:m_F-all} do
not need to be computed as they have the same eigenenergies and the
twofold degenerate eigenstates just differ by time-reversal symmetry.
The total number $N$ of basis states and thus the dimension of the
generalized eigenvalue problem~\eqref{eq:generalized_ev_problem} is
given by the product of the number of B-splines for the $\rho$,
$z_{\mathrm{e}}$, and $z_{\mathrm{h}}$ coordinates and the number of
angular momentum and spin states $|m,J,m_J\rangle$, and was up to
$N=26\times 15\times 15\times 18=105\,300$ in our calculations.

\begin{figure*}
\includegraphics[width=\textwidth]{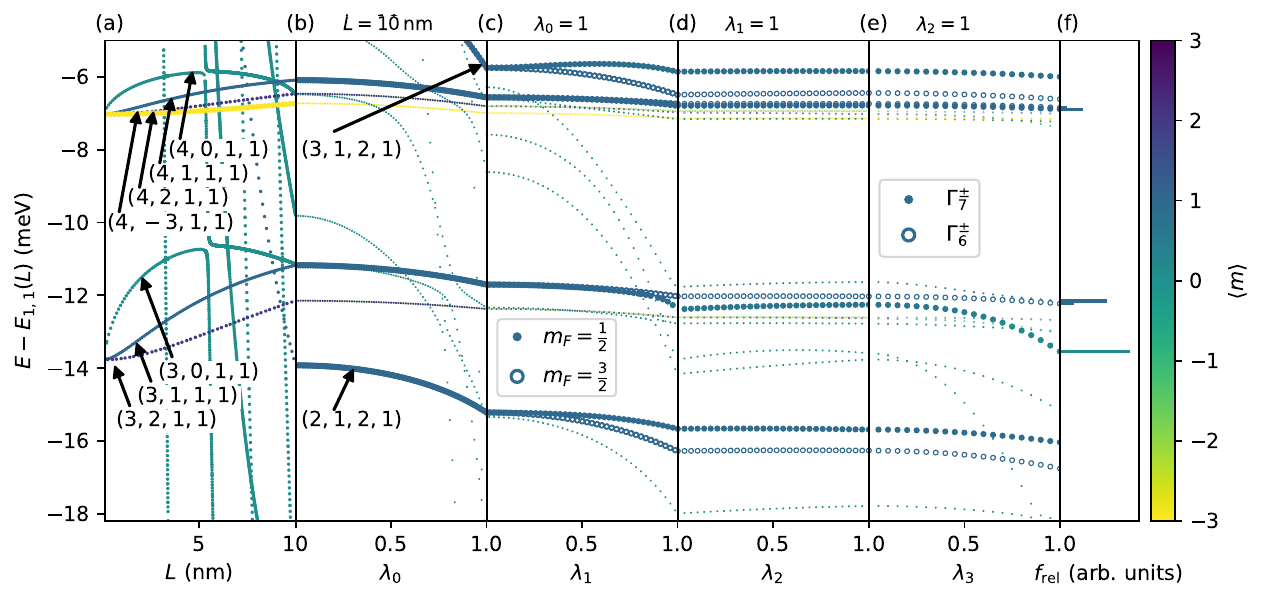}
\caption{Exciton spectra in a cuprous oxide QW.  In (a) the QW width
  is increased from $L=0$ to $10\,$nm for excitons described by the
  hydrogenlike part of the Hamiltonian given by the first two terms in
  Eq.~\eqref{eq:H_lambda}.  In (b)-(e) the nondiagonal coupling terms
  are successively switched on using the parameters
  $\lambda_{0,1,2,3}\in [0,1]$ in the Hamiltonian~\eqref{eq:H_lambda}
  (see text).  The photoabsorption spectrum for transitions with
  circularly polarized light and considering the complete structure of
  the valence band is given in (f).  Some states are labeled by
  approximate quantum numbers $(n,m,N_{\mathrm{e}},N_{\mathrm{h}})$.
  The colors of symbols or lines indicate the expectation value
  $\langle m\rangle$ of the angular momentum of the exciton state.}
\label{fig:lambda_plot}
\end{figure*}
To compute oscillator strengths we use the eigenstates, obtained by
solving the generalized  eigenvalue problem.
The solution yields the wave function for each state in the
coordinates $\rho$, $z_\mathrm{e}$, and $z_\mathrm{h}$. For evaluating
Eqs.~\eqref{eq:transition_amplitudes}, the wave functions have to be
expressed in relative coordinates in the $z$ direction. As we only
need the wave function at $z=0$, i.e., at $ z_\mathrm{e}=z_\mathrm{h}=Z$,
the transformation is straightforward and given by
\begin{equation}
  \psi(\rho,z\mathord{=}0,Z) =\psi(\rho,z_\mathrm{e}\mathord{=}Z,z_\mathrm{h}\mathord{=}Z).
\end{equation}
The derivative  with respect to $z$ is calculated
analytically by differentiating the B-spline basis functions according
to Eq.~\eqref{eq:recursion_bsplines}.
It acts on a product of basis states as
\begin{equation}
\partial_z B_j^{n}(z_{\mathrm e})\,B_k^{n}(z_{\mathrm h}) = \partial_{z_\mathrm{e}} B_j^{n}(z_{\mathrm e})\,B_k^{n}(z_{\mathrm h})-B_j^{n}(z_{\mathrm e})\,\partial_{z_\mathrm{h}} B_k^{n}(z_{\mathrm h}).
\end{equation}
By contrast, the derivative with respect to $\rho$ at $\rho=0$ is
computed using the approximation
$\partial_\rho\psi(\rho,z\mathord{=}0,Z) \approx \frac{1}{\rho}\psi(\rho,z\mathord{=}0,Z)$,
which is extrapolated, analogously to Ref.~\cite{Kuehner2025}, to $\rho=0$.
In Eq.~\eqref{eq:relativeOscillatorStrength_integral}
the contributions of both twofold degenerate states to the
relative oscillator strength $f_{\mathrm{rel}}$ are considered.
The integrands are evaluated at 100 positions on an equidistant grid,
and the integrals are solved numerically using \texttt{numpy.trapz}~\cite{numpy}.
Note that the most time consuming part is the diagonalization of the
generalized eigenvalue problem~\eqref{eq:generalized_ev_problem}.

\section{Results and discussion}
\label{sec:results}
The complex structure of the cuprous oxide valence bands has a
significant effect on excitons in the QW.
To observe the shifts of energy levels due to the impact of the
different terms in the Hamiltonian~\eqref{eq:H_final}, we apply the
following strategy.
We start by taking into account the diagonal part of the Hamiltonian,
\begin{equation}\label{eq:H_hyd}
  H_{\mathrm{hyd}} = H_{\mathrm{diag}} + H_{\Delta m=0}\delta_{J'J}\, ,
\end{equation}
representing the hydrogenlike two-band model of excitons
in a QW, and then switch on, term by term, the parts in the
Hamiltonian, which are nondiagonal with respect to
the angular momentum and spin quantum numbers of the basis.
To this aim, we modify the Hamiltonian~\eqref{eq:H_final} to the form
\begin{align}
  H &= H_{\mathrm{diag}} + H_{\Delta m=0}\delta_{J'J}
  + {\lambda_0} H_{\Delta m=0}(1-\delta_{J'J})\nonumber \\
  &+ {\lambda_1} H_{\Delta m=1}
  + {\lambda_2} H_{\Delta m=2} + {\lambda_3} H_{\Delta m_F=4}
\label{eq:H_lambda}
\end{align}
by introducing the auxiliary parameters $\lambda_{0,1,2,3}\in [0,1]$,
which can be used to successively switch on the individual
nondiagonal terms.
Note that $H_{\Delta m=0}$ in Eq.~\eqref{eq:H_Dm=0} has both,
nonzero diagonal and nondiagonal terms in the angular momentum and
spin basis $|m,J,m_J\rangle$. In Eq.~\eqref{eq:H_lambda}, the nonzero
diagonal terms of $H_{\Delta m=0}$ are considered in the hydrogenlike
model, described by the first and second term, and the nondiagonal
terms, which couple $J=1/2$ and $J=3/2$ states of the yellow and
green exciton series, respectively, are switched on by increasing
$\lambda_0$ from $0$ to $1$.

For an energy range of the lowest, yellow exciton series a sequence
of spectra obtained by diagonalization of the
Hamiltonian~\eqref{eq:H_lambda} is presented in Fig.~\ref{fig:lambda_plot}.
The figure allows for a detailed discussion of the impact of the
valence band on excitons in a cuprous oxide QW of $L=10\,$nm width.
We go through Fig.~\ref{fig:lambda_plot} from left to right.
Panel (a) shows spectra of the hydrogenlike two-band model,
Eq.~\eqref{eq:H_hyd} for excitons in a QW of width $L$ in the range
from $L=0$ to $L=10\,$nm.
The energies are given with respect to the first scattering
threshold $E_{1,1}(L)$, given by the sum of the lowest
quantum-confinement energies, see Eq.~\eqref{eq:E_QW}.
Note that in the limit $L\to 0$ both the exciton energies $E$
and the threshold energy $E_{1,1}(L)$ diverge, however, the energy
differences $E-E_{1,1}(L)$ remains finite.
This limit refers to the analytically solvable two-dimensional Coulomb
problem~\cite{Ziemkiewicz2021a,Ziemkiewicz2021b}, where states with
the same principal
quantum number $n$, but different angular momentum quantum numbers $m$
are degenerate, as can be clearly seen in Fig.~\ref{fig:lambda_plot}.
The degeneracy is lifted, when $L$ is increased, in the transition
region from strong to weak confinement~\cite{Belov2024}.
Some of the states in panel (a) are marked by quantum numbers
$(n,m,N_{\mathrm{e}},N_{\mathrm{h}})$, with $m$ being an exact quantum
number due to the rotational symmetry of the hydrogenlike problem.
The colors of the states indicate the expectation values $\langle m\rangle$
of the angular momentum operator throughout Fig.~\ref{fig:lambda_plot}.

In panels (b)--(f) the QW width is fixed to $L=10\,$nm
to study the impact of the complex valence band on the excitons at
such a QW width.
In panel (b) the nondiagonal terms of $H_{\Delta m=0}$, coupling
the yellow ($J=1/2$) and green ($J=3/2$) excitons, are switched on by
increasing $\lambda_0$ from $0$ to $1$.
The coupling between yellow and green excitons has a strong impact on
states with angular quantum number $m=0$, and smaller effect on states
with $m\ne 0$, the energies of those are moderately lowered.
Note that for the states in this panel the rotational symmetry around
the $z$ axis is still valid, and thus $m$ and $m_J$ are exact quantum
numbers.
As the eigenenergies do not depend on the sign of these quantum
numbers, states with $m\ne 0$ are fourfold degenerate.

In panels (c) and (d) of Fig.~\ref{fig:lambda_plot} the
nondiagonal terms $H_{\Delta m=1}$ and $H_{\Delta m=2}$ in the
Hamiltonian~\eqref{eq:H_lambda} are successively switched on.
These coupling terms break the rotational symmetry around
the $z$ axis in a way that $m$ is not a good quantum number any longer,
but the quantum number $m_F=m+m_J$ for the $z$ component of the total
angular momentum $\boldsymbol{F}=\boldsymbol{L}+\boldsymbol{J}$ still is.
As an important consequence, the degeneracy of, e.g., some states with
$m_F=1/2$ and $m_F=3/2$, marked by solid and open symbols, respectively,
is reduced in panels (c) and (d) of Fig.~\ref{fig:lambda_plot}
from fourfold to twofold degeneracy.

Finally, in panel (e) the nondiagonal terms $H_{\Delta m_F=4}$ in
the Hamiltonian~\eqref{eq:H_lambda}, which couple basis states with
$\Delta m=\pm 2$ and $\Delta m_F=\pm 4$, are switched on.
In this panel, $m_F$ is not a good quantum number any longer, and states
can only be classified as belonging to one of the four irreducible
subspaces $\Gamma_6^\pm$ and $\Gamma_7^\pm$, as discussed in
Sec.~\ref{sec:symmetries}.
The eigenenergies are shifted due to the coupling between different
$m_F$ states, however, there is no further lifting of degeneracies,
i.e., all states remain twofold degenerate.
The rightmost spectrum in Fig.~\ref{fig:lambda_plot}(e) is the one with
$\lambda_{0,1,2,3}=1$, where the full impact of the valence band
as described by the Hamiltonian~\eqref{eq:H_final} is taken into account.
The relative oscillator strengths corresponding to this spectrum are shown in
Fig.~\ref{fig:lambda_plot}(f).
Note that the spectrum significantly differs from that of the
hydrogenlike model shown at QW width $L=10\,$nm in Fig.~\ref{fig:lambda_plot}(a).
This clearly illustrates that a precise description of exciton states in QWs
requires taking into account the complex structure of the valence band
to accurately determine all splittings of energy levels.

\begin{figure}
\includegraphics[width=\columnwidth]{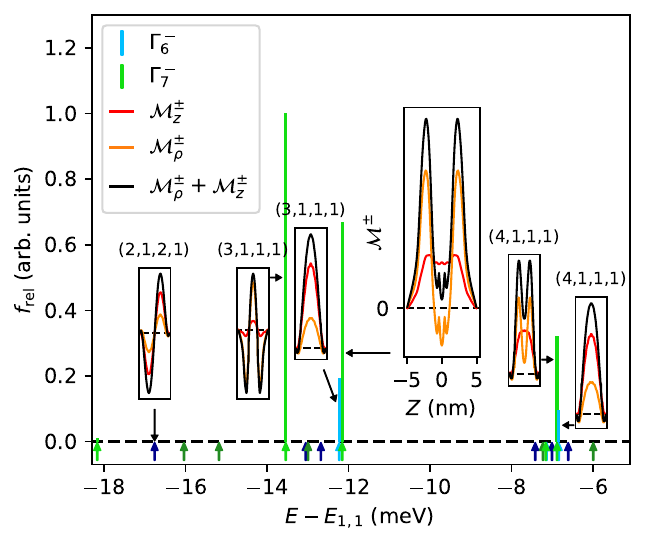}
\caption{Photoabsorption spectrum of a cuprous oxide quantum well of
  width $L=10\,$nm for transitions with circularly polarized light
  ($\hat{\bm e}_x\pm i\hat{\bm e}_y$).  The relative oscillator strengths
  $f_{\mathrm{rel}}$ of parity allowed transitions to states in the subspaces
  $\Gamma_6^-$ and  $\Gamma_7^-$ are represented by vertical light blue and
  green  bars, respectively.  Parity forbidden even states in the
  subspaces $\Gamma_6^+$ and  $\Gamma_7^+$ are marked by dark blue and
  dark green arrows.  The insets show the transition amplitude density
  $\mathcal{M}_\rho^\pm(Z)$ (orange lines) and $\mathcal{M}_z^\pm(Z)$
  (red lines) given in Eqs.~\eqref{eq:transition_amplitudes1} and
  \eqref{eq:transition_amplitudes2}, respectively, together with
  their sums (black lines).  Some states are marked by approximate
  quantum numbers $(n,m,N_{\mathrm{e}},N_{\mathrm{h}})$.}
\label{fig:osc_circ_plus}
\end{figure}
For a more detailed discussion, the photoabsorption spectrum of a
cuprous oxide QW of width $L=10\,$nm is shown in
Fig.~\ref{fig:osc_circ_plus}.
Excitons are excited by a laser beam oriented perpendicular to the QW
plane with left- or right-handed ($\hat{\bm e}_x\pm i\hat{\bm e}_y$)
circularly polarized light.
Note that due to the selection rules discussed in Sec.~\ref{sec:selection_rules}
only states in the irreducible subspaces $\Gamma_6^-$ and $\Gamma_7^-$
can be excited.
Their relative oscillator strengths $f_{\mathrm{rel}}$
are computed
by evaluation of Eq.~\eqref{eq:relativeOscillatorStrength_integral}
and indicated in Fig.~\ref{fig:osc_circ_plus} by vertical light
green and light blue bars, respectively.
Parity-forbidden even states in the subspaces $\Gamma_6^+$ and
$\Gamma_7^+$ are marked by dark blue and dark green arrows.
The insets show the transition amplitude density $\mathcal{M}_\rho^\pm(Z)$
(orange lines) and $\mathcal{M}_z^\pm(Z)$ (red lines) given in
Eqs.~\eqref{eq:transition_amplitudes1} and \eqref{eq:transition_amplitudes2},
respectively.
As can be clearly seen, they are even functions in $Z$ for the parity
allowed and odd functions for the parity forbidden states.
It should be noted that, contrary to the simple two-band model
discussed in Ref.~\cite{Kuehner2025}, both densities
$\mathcal{M}_\rho^\pm(Z)$ and $\mathcal{M}_z^\pm(Z)$ are nonzero and
constructively or destructively interfere to the sums shown as black
lines in the insets of Fig.~\ref{fig:osc_circ_plus}.

\begin{table}
\caption{Numerical values for selected states of the exciton spectrum
  at QW width $L=10\,$nm considering the complex band structure
  described by the Hamiltonian~\eqref{eq:H_final}. The energy eigenvalues $E$ are
  compared to those of the hydrogenlike model $H_{\mathrm{hyd}}$ given
  in Eq.~\eqref{eq:H_hyd}. Approximate quantum numbers
  $(n,m,N_{\mathrm{e}},N_{\mathrm{h}})$ are assigned where it was
  uniquely possible. All states are characterized by their irreducible
  representations. Oscillator strengths $f_{\mathrm{rel}}$ are given
  relative to the strongest line.}
\vspace{1ex}
\begin{tabular}{rrrrcccc}
  $E\,$(meV) & $E_{\mathrm{hyd}}\,$(meV) & $~~n$ & $m$ &
  $N_{\mathrm{e}}$ & $N_{\mathrm{h}}$ & sym & $f_{\mathrm{rel}}$ \\[0.5ex]
  \hline
$-112.95986$         & $-97.28582$          &1   &0  &1  &1     &$\Gamma_7^+$      &0 \\
$-65.16197 $         & $-24.64831$          &2   &0  &1  &1     &$\Gamma_7^+$      & 0 \\
$-27.80125 $         & $-26.25192$          &2   &1  &1  &1     &$\Gamma_6^-$      & 0.39413\\
$-27.37152 $         & $-26.25192$          &2   &1  &1  &1     &$\Gamma_7^-$      & 0.00251\\
$-18.16818$ & $-9.81564$	 & &&&            &$\Gamma_{7}^{-}$& 0.01186 \\
$-16.76044$ & $-13.92396$ &$2$ & $1$ & $2$ & $1$  &$\Gamma_{6}^{+}$& 0 \\
$-16.04019$ & $-13.92396$ &$2$ & 1 & 2 & $1$  &$\Gamma_{7}^{+}$& 0 \\
$-15.18094$ & $-4.53941$ & &&&         &$\Gamma_{7}^{+}$& 0 \\
$-13.54263$ & $-11.17306$ &$3$ & 1 & 1 & $1$  &$\Gamma_{7}^{-}$& $1.00000$ \\
$-13.05286$ & $-11.14293$ &$3$ & 2 & 1 & $1$  &$\Gamma_{6}^{+}$& 0 \\
$-12.99400$ & $-6.48151$	 &$4$ & 0 & 1 & $1$  &$\Gamma_{7}^{+}$& 0 \\
$-12.68154$ & $-11.14293$ &$3$ & $-2$ & 1 & $1$  &$\Gamma_{6}^{+}$& 0 \\
$-12.22872$ & $-11.17306$ &$3$ & 1 & 1 & $1$  &$\Gamma_{6}^{-}$& $0.19371$ \\
$-12.15775$ & $      $	 &&&&         &$\Gamma_{7}^{-}$& $0.66918$ \\
$-7.42073$  & $-6.46143$	 &$4$ & 2 & 1 & $1$  &$\Gamma_{6}^{+}$& 0 \\
$-7.23267$  & $-3.34728$	 &&&&         &$\Gamma_{7}^{+}$& 0 \\
$-7.16253$  & $-6.72711$	 &$4$ & $-3$ & 1 & $1$ &$\Gamma_{7}^{-}$& $0.00012$ \\
$-7.14464$  & $-6.72711$	 &$4$ & $-3$ & 1 & $1$ &$\Gamma_{6}^{-}$& 0.00002\\
$-7.01147$  & $-6.46143$	 &$4$ & $-2$ & 1 & $1$ &$\Gamma_{6}^{+}$& 0 \\
$-6.88742$  & $-6.08855$	 &$4$ & 1 & 1 & $1$  &$\Gamma_{7}^{-}$& $0.32152$ \\
$-6.84615$  & $-6.08855$	 &$4$ & 1 & 1 & $1$  &$\Gamma_{6}^{-}$& 0.09737 \\
$-6.61281$  & $-2.08170$	 &$3$ & 1 & 2 & $1$  &$\Gamma_{6}^{+}$& 0\\
$-5.99680$  & $-2.08170$	 &$3$ & 1 & 2 & $1$  &$\Gamma_{7}^{+}$& 0
\end{tabular}
\label{tab:spectrum}
\end{table}
For convenience, selected energy eigenvalues $E$ of the Hamiltonian~\eqref{eq:H_final}
at QW width $L=10\,$nm corresponding to the right-most spectrum in
Fig.~\ref{fig:lambda_plot}(e) and the spectrum in
Fig.~\ref{fig:osc_circ_plus} are given in Table~\ref{tab:spectrum}.
They are compared to the energies $E_{\mathrm{hyd}}$ of the hydrogenlike model,
i.e., the right-most spectrum shown in Fig.~\ref{fig:lambda_plot}(a).
Where possible, approximate quantum numbers
$(n,m,N_{\mathrm{e}},N_{\mathrm{h}})$ are assigned to the states.
The irreducible representations for the states calculated using the
complete valence band and the relative oscillator strengths
$f_{\mathrm{rel}}$ shown in Figs.~\ref{fig:lambda_plot}(f) and
\ref{fig:osc_circ_plus} are also indicated in Table~\ref{tab:spectrum}.
\begin{figure}
  \includegraphics[width=0.98\columnwidth]{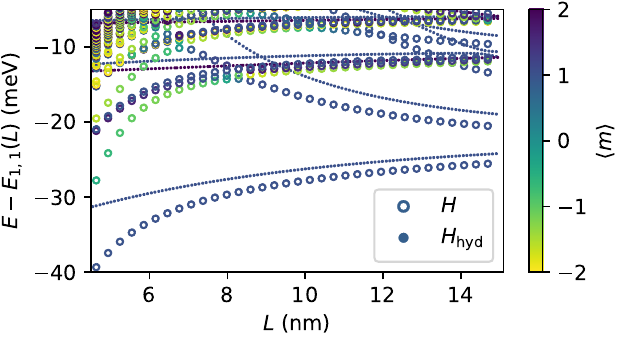}
  \caption{Comparison of the exciton spectra of the
    $\Gamma_{6}^{\pm}$ states in a cuprous oxide quantum well as
    a function of the well width $L$. Results obtained from the
    full Hamiltonian $H$ in Eq.~\eqref{eq:H_final} and the 
    hydrogenlike model based on the Hamiltonian $H_{\mathrm{hyd}}$
    given in Eq.~\eqref{eq:H_hyd} are presented by open and closed
    symbols, respectively.  The threshold energy $E_{1,1}(L)$ has been
    subtracted from all energy eigenvalues.}
\label{fig:width_a}
\end{figure}

So far, we have discussed the impact of the valence band on exciton
spectra in a QW with fixed width $L=10\,$nm.
In Fig.~\ref{fig:width_a} we finally compare the dependence of the
energy eigenvalues of the full Hamiltonian~\eqref{eq:H_final} (open
symbols) and of the hydrogenlike model in Eq.~\eqref{eq:H_hyd} (closed
symbols) on the QW width $L$.
The change of energy levels during the crossover from the 2D limit
(strong confinement regime) to the bulk semiconductor (weak
confinement) in the range $5\,\mathrm{nm}\le L\le 15\,\mathrm{nm}$ can
be observed.
Since the Bohr radius of the lowest calculated exciton state is about
4~nm, the QW width of 15 nm can be considered for this state as a weak
confinement regime.
States obtained within the simplified hydrogenlike model exhibit only
moderate variations in their separation from the threshold level
$E_{1,1}(L)$. By contrast, due to strong state-mixing effects, the
levels calculated using the more complex model show a pronounced
sensitivity to changes in the QW width, particularly in the regime of
small $L$.
Energy levels for $L<5\,$ nm are not shown because for such widths the
Luttinger-Kohn model with the same material parameters as for the bulk
is very approximate (see Sec.~\ref{sec:conclusion}).

\section{Conclusion and outlook}
\label{sec:conclusion}
We have studied the impact of the valence band structure
on the energies and level splittings of yellow excitons in
cuprous oxide QWs.
Based on the Luttinger-Kohn model, we derived the complete
Hamiltonian, taking into account the full complex valence band
structure.
We calculated the exciton energies by expanding the wave function over
a multidimensional basis of angular and spin eigenstates and the
B-splines for the coordinate part of the wave function.
This allowed us to observe the breaking of the rotational symmetry,
leading to the quantum number $m$ no longer being a good
quantum number.
The step-by-step inclusion of the nondiagonal coupling terms of the
complex valence band Hamiltonian clearly demonstrated shifts and
splittings of the energy levels.
We also computed the relative oscillator strengths for the excitonic
transitions in QWs with circularly polarized light.

In this article, we have considered the complete valence band structure
but have still used some approximations and simplifications, which
shall be addressed in future work.
In the Hamiltonian~\eqref{eq:H_cartesian} we have used the same
Luttinger parameters as in the bulk material~\cite{Schoene16}.
This may be a good approximation for sufficiently broad QWs
($L\gtrsim 5$~nm),
however, the Luttinger parameters probably need to be adjusted for
narrow QWs based on density functional theory (DFT)
calculations~\cite{French2009} of the electronic band structure of the
thin films~\cite{Chaves2020}.
In our calculations we have also assumed the same dielectric constant
$\varepsilon$ for both the crystal and the substrate.
The dielectric contrast of a more realistic setup can be
taken into account via the Rytova-Keldysh potential~\cite{Rytova,Keldysh}
similar to what has been done for the hydrogenlike model of excitons in
QWs~\cite{Belov2024}.
Furthermore, we have restricted our computations to
a configuration in which the QW is aligned perpendicular to the
$[001]$ axis of the crystal, and excitons are invariant under symmetry
transformations of the $D_{4\mathrm{h}}$ symmetry group.
A more general orientation between the QW and the crystal can be
considered, however, this will further reduce the symmetries of the problem.
For the computation of spectra and oscillator strengths we have
focused on bound states of the yellow exciton series.
In future work, this can be generalized to yellow and green resonance
states above threshold by application of the complex-coordinate-rotation
method as in Ref.~\cite{Scheuler2024}.
This will also enable the search for BICs~\cite{Aslanidis2025}.
The obtained wave functions will make it possible to estimate the
interband dipole transition matrix elements to investigate the usage
of the confined Rydberg excitons for quantum sensing.
The consideration of all these effects and radiative properties will
certainly allow for detailed comparisons with future experimental spectra.


\acknowledgments
Stefan Scheel and Pavel A.\ Belov acknowledge financial support by the
National Research Council Canada, grant No.~QSP-129-1.


\appendix
\section{B-splines}
\label{app:B-splines}
B-splines of higher orders are an effective tool for discretizing and
solving partial differential equations.
The unknown function is expanded over a basis of B-splines
$B^{k}_{i}(x)$, $i=1,\ldots,\mathcal{N}$, which are piecewise polynomials of
degree $k-1$.
Given the predefined series of the service nodes $t_{i}$, each
B-spline $B^{k}_{i}(x)$ of order $k$ is defined on the interval
$x\in [t_{i},t_{i+k}]$.
Values of the B-splines $B^{k}_{i}(x)$ and their derivatives at a
given point $x$ can be calculated by recursion formulas~\cite{Bachau2001}
\begin{align}\nonumber
B^{k}_{i}(x) &= \frac{x-t_{i}}{t_{i+k-1}-t_{i}} B^{k-1}_{i}(x)+
\frac{t_{i+k}-x}{t_{i+k}-t_{i+1}} B^{k-1}_{i+1}(x)\, ,\\
\frac{d B^{k}_{i}(x)}{dx} &= \frac{k-1}{t_{i+k-1}-t_{i}} B^{k-1}_{i}(x)-
\frac{k-1}{t_{i+k}-t_{i+1}} B^{k-1}_{i+1}(x) \, .\label{eq:recursion_bsplines}
\end{align}
The expression for the second derivative can be derived from the above two equations.

B-splines have some important characteristics.
First, a B-spline of order $k$ on an equidistant grid approximates an
analytical function with accuracy of $h^{k}$, where $h$ is the
step size of the grid.
Thus, higher-order B-splines give an accurate solution even for a
relatively small number of nodes, which is particularly important for
multi-dimensional problems.
Second, one can choose nodes non-equidistantly and add service nodes
at the boundaries in such a way as to have some B-splines equal to 1
at the boundaries, while all other B-splines are exactly zero there.
Then, for example, zero boundary conditions can be easily implemented
by removing the B-splines which are nonzero at the boundaries.
Third, the B-splines are nonorthogonal functions, and thus the problem
turns into a generalized eigenvalue problem.
However, the B-spline functions have minimal support, i.e., each B-spline
vanishes, $B^{k}_{i}(x) = 0$, for $x \notin [t_{i},t_{i+k}]$, which
significantly reduces the number of integrations to calculate matrix
elements.
This leads to a sparse structure of the matrices of the generalized
eigenvalue problem~\eqref{eq:generalized_ev_problem}.

\section{Matrix elements of momentum operators}
\label{app:momentum_operators}
In polar coordinates $(\rho,\phi)$ the creation and annihilation
operators $p_\pm$ are given as
\begin{align}
  p_\pm=\frac{\hbar}{i}\exp(\pm i\phi)
  \left [ \partial_\rho \mp\frac{1}{\rho} \left(\frac{L_z}{\hbar}\pm \frac{1}{2}\right)\right ]
\label{eq:p_pm}
\end{align}
with $p_+ = p_-^\dagger$.
Using Eq.~\eqref{eq:p_pm} and the commutator
$[L_z/\hbar,\exp(\pm i\phi)]=\pm1$
the squared operators $p_\pm^2$ are obtained as
\begin{align}
  p_\pm^2 =&-\hbar^2\exp(\pm 2i\phi)  \left
           [\partial_\rho^2\mp \frac{2}{\rho}\partial_\rho\left(\frac{L_z}{\hbar}\pm 1\right )
           \right. \nonumber\\
  &+ \left. \frac{1}{\rho^2}
     \left(\frac{L_z}{\hbar}\pm\frac{5}{2}\right)\left(\frac{L_z}{\hbar}\pm\frac{1}{2}\right)
           \right] \, .
\label{eq:p2_pm}
\end{align}
The operator $p_+p_-$ reads
\begin{align}
  p_+p_- = -\hbar^2\partial_\rho^2 + \frac{1}{\rho^2}\left(L_z^2-\frac{\hbar^2}{4}\right)
  \, ,
\end{align}
and thus is diagonal with respect to the angular momentum functions
$\exp(im\phi)$.
For functions given as a product of a B-spline over the $\rho$
coordinate and an angular function $\exp(im\phi)$, it is now
straightforward to evaluate matrix elements of the operators
given above.
The $\phi$ integrals are trivially solved analytically, yielding
selection rules for non-zero matrix elements between states with
quantum numbers $m$ and $m'$.
The B-spline functions as well as their derivatives are piecewise
polynomials, and can be easily integrated.
Note that due to the boundary condition that the B-splines vanish at
$\rho=0$, the product of two B-splines can be divided by $\rho$ or
$\rho^2$ without the occurrence of any singularity at the origin.

Using B-splines over the coordinates $z_{\mathrm{e}}$ and $z_{\mathrm{h}}$,
matrix elements of the operators
$p_{\mathrm{e}z}^2=-\hbar^2\partial_{\mathrm{e}z}^2$,
$p_{\mathrm{h}z}=(\hbar/i)\partial_{\mathrm{h}z}$, and
$p_{\mathrm{h}z}^2=-\hbar^2\partial_{\mathrm{h}z}^2$
are straightforwardly computed by integration of products of B-splines
and their derivatives.

\section{Matrix elements of spin operators}
\label{app:spin_operators}
The matrix elements of the quasispin and hole spin operators $\bm{I}$
and $\bm{S}_{\mathrm{h}}$ are computed in the six-dimensional basis
$|J,m_J\rangle$ of the spin operator $\bm{J}=\bm{I}+\bm{S}_{\mathrm{h}}$,
which can also be expressed, using Clebsch-Gordan coefficients,
in the basis $|I,m_I;S_{\mathrm{h}},m_{S_{\mathrm{h}}}\rangle$, i.e.
\allowdisplaybreaks[2]
\begin{align*}
  |\tfrac{3}{2},\tfrac{3}{2}\rangle  &=\ket{1,1;\tfrac{1}{2},\tfrac{1}{2}}\, , \\
  |\tfrac{3}{2},\tfrac{1}{2}\rangle &=\sqrt{\tfrac{2}{3}}\ket{1,1;\tfrac{1}{2},-\tfrac{1}{2}}-\sqrt{\tfrac{1}{3}}\ket{1,0;\tfrac{1}{2},\tfrac{1}{2}}\, , \\
  |\tfrac{1}{2},\tfrac{1}{2}\rangle &=\sqrt{\tfrac{1}{3}}\ket{1,1;\tfrac{1}{2},-\tfrac{1}{2 }}+\sqrt{\tfrac{2}{3}}\ket{1,0;\tfrac{1}{2},\tfrac{1}{2}} \, , \\
  |\tfrac{1}{2}, -\tfrac{1}{2}\rangle &=  -\sqrt{\tfrac{2}{3}}\ket{1,-1;\tfrac{1}{2},\tfrac{1}{2}}+\sqrt{\tfrac{1}{3}}\ket{1,0;\tfrac{1}{2},-\tfrac{1}{2}}\, , \\
  |\tfrac{3}{2}, -\tfrac{1}{2}\rangle &=\sqrt{\tfrac{1}{3}}\ket{1,-1;\tfrac{1}{2},\tfrac{1}{2 }}+\sqrt{\tfrac{2}{3}}\ket{1,0;\tfrac{1}{2},-\tfrac{1}{2}}\, , \\
  |\tfrac{3}{2}, -\tfrac{3}{2}\rangle &= \ket{1,-1;\tfrac{1}{2},-\tfrac{1}{2}}\, .
\end{align*}
With these basis states the matrix representations of the spin
operators are obtained as
\allowdisplaybreaks[2]
\begin{align}
  I_+ &= \hbar
  \begin{pmatrix}
  	0 & -\tfrac{\sqrt{6}}{3} & \tfrac{2\sqrt{3}}{3} & 0 & 0 & 0 \\
  	0 & 0 & 0 & \tfrac{4}{3} & \tfrac{\sqrt{2}}{3} & 0 \\
  	0 & 0 & 0 & -\tfrac{\sqrt{2}}{3} & \tfrac{4}{3} & 0 \\
  	0 & 0 & 0 & 0 & 0 & \tfrac{\sqrt{6}}{3} \\
  	0 & 0 & 0 & 0 & 0 & \tfrac{2\sqrt{3}}{3} \\
  	0 & 0 & 0 & 0 & 0 & 0
  \end{pmatrix} \, , \\[1ex]
  I_3 &=\hbar
  \begin{pmatrix}
  	1 & 0 & 0 & 0 & 0 & 0 \\
  	0 & \tfrac{2}{3} & \tfrac{\sqrt{2}}{3} & 0 & 0 & 0 \\
  	0 & \tfrac{\sqrt{2}}{3} & \tfrac{1}{3} & 0 & 0 & 0 \\
  	0 & 0 & 0 & -\tfrac{2}{3} & \tfrac{\sqrt{2}}{3} & 0 \\
  	0 & 0 & 0 & \tfrac{\sqrt{2}}{3} & -\tfrac{1}{3} & 0 \\
  	0 & 0 & 0 & 0 & 0 & -1
  \end{pmatrix} \, , \\[1ex]
  S_{\mathrm{h}+} &= \hbar
  \begin{pmatrix}
  	0 & \tfrac{\sqrt{6}}{3} & \tfrac{\sqrt{3}}{3} & 0 & 0 & 0 \\
  	0 & 0 & 0 & -\tfrac{1}{3} & -\tfrac{\sqrt{2}}{3} & 0 \\
  	0 & 0 & 0 & \tfrac{\sqrt{2}}{3} & \tfrac{2}{3} & 0 \\
  	0 & 0 & 0 & 0 & 0 & -\tfrac{\sqrt{6}}{3} \\
  	0 & 0 & 0 & 0 & 0 & \tfrac{\sqrt{3}}{3} \\
  	0 & 0 & 0 & 0 & 0 & 0
  \end{pmatrix} \, , \\[1ex]
  S_{\mathrm{h}3} &= \hbar
  \begin{pmatrix}
  	\tfrac{1}{2} & 0 & 0 & 0 & 0 & 0 \\
  	0 & -\tfrac{1}{6} & -\tfrac{\sqrt{2}}{3} & 0 & 0 & 0 \\
  	0 & -\tfrac{\sqrt{2}}{3} & \tfrac{1}{6} & 0 & 0 & 0 \\
  	0 & 0 & 0 & \tfrac{1}{6} & -\tfrac{\sqrt{2}}{3} & 0 \\
  	0 & 0 & 0 & -\tfrac{\sqrt{2}}{3} & -\tfrac{1}{6} & 0 \\
  	0 & 0 & 0 & 0 & 0 & -\tfrac{1}{2}
  \end{pmatrix} \, .
\end{align}
The matrix elements of the annihilation operators are given by the
conjugate transpose of the creation operator matrices, i.e.,
$I_-=I_+^\dagger$ and $S_{\mathrm{h}-}=S_{\mathrm{h}+}^\dagger$.
The matrix elements of products of spin operators can be derived by
multiplication of the corresponding spin matrices.
Using $\bm{I}\cdot\bm{S}_{\mathrm{h}}=\frac{1}{2}(\bm{J}^2-\bm{I}^2-\bm{S}_{\mathrm{h}}^2)$,
its matrix elements are
\begin{align}
  \langle J'm_J'|\bm{I}\cdot\bm{S}_{\mathrm{h}}|Jm_J\rangle
  = \frac{1}{2}\left(J(J+1)-\frac{11}{4}\right) \delta_{J'J}\delta_{m_J'm_J} \, .
\end{align}

\bibliography{paper}

\end{document}